\documentclass[%
 aip,
 amsmath,amssymb,
 reprint,%
]{revtex4-1}

\usepackage{graphicx}% Include figure files
\usepackage{dcolumn}% Align table columns on decimal point
\usepackage{bm}% bold math
\usepackage{xcolor}
\usepackage[utf8]{inputenc}
\usepackage[T1]{fontenc}
\usepackage{mathptmx}

\begin{document}

\preprint{AIP/123-QED}

\title{Two-dimensional optical chimera states in an array of coupled waveguide resonators}
% Force line breaks with \\

\author{M.G. Clerc}
\affiliation{Departamento de F\'isica and Millennium Institute for Research in Optics, Facultad de Ciencias F\'isicas y Matem\'aticas, Universidad de Chile, Casilla 487-3, Santiago, Chile}%Lines break automatically or can be forced with \\
\author{S. Coulibaly}%
% \email{Second.Author@institution.edu.}
\affiliation{Universit\'e de Lille, CNRS, UMR 8523-PhLAM-Physique des Lasers Atomes et Mol\'ecules, F-59000 Lille, France}%

\author{ M.A. Ferr\'e}
% \homepage{http://www.Second.institution.edu/~Charlie.Author.}
\affiliation{%
Departamento de F\'isica and Millennium Institute for Research in Optics, Facultad de Ciencias F\'isicas y Matem\'aticas, Universidad de Chile, Casilla 487-3, Santiago, Chile%\\This line break forced% with \\
}%
\author{M. Tlidi}
\affiliation{Facult\'e des Sciences, Universit\'e Libre de Bruxelles (U.L.B), CP 231, Campus Plaine, B-1050 Bruxelles, Belgium}

\date{\today}% It is always \today, today,

\begin{abstract}
Two-dimensional arrays of coupled waveguides or coupled microcavities allow to confine 
and manipulate light. Based on a paradigmatic envelope equation, we show that these devices, 
subject to a coherent optical injection, support coexistence between a coherent and incoherent emission. In this regime, we show that 
two-dimensional chimera state can be generated.
Depending on initial conditions, the system exhibits a family of two-dimensional chimera 
states and interaction between them. We characterize these  
two-dimensional structures by computing their Lyapunov spectrum, 
and Yorke-Kaplan dimension. Finally, we show that two-dimensional chimera states are of spatiotemporal chaotic nature.
\end{abstract}

\maketitle

\begin{quotation}
One-dimensional nonlinear coupled microcavities exhibit a rich spatiotemporal dynamics. 
In particular, these coupled microcavities have fully synchronized or incoherent light emission 
of a spatiotemporal chaotic nature. Also, depending on the initial conditions, 
these devices show coexistence between desynchronized and synchronized domains, 
often called {\it optical chimera states}. In this contribution, we show evidence {of} optical chimeras 
in a two-dimensional array of coupled waveguide resonators. 
Due to the additional degrees of freedom, the smaller localized solutions exhibit a chaotic 
spatiotemporal evolution---which is not the case of the one-dimensional counterpart. 
Lyapunov spectrum and Yorke-Kaplan dimensions are calculated to characterize these {intriguing} localized states.
\end{quotation}

\section{Introduction}

  A two-dimensional array of coupled waveguides or coupled microcavities consists of nonlinear discrete structures  \cite{Joseph1988}. This configuration appears not only in photonics but also in a large variety of systems such as {biological systems}
%biology 
\cite{Davydov}, condensed matter physics  \cite{Su1979}, and Bose-Einstein condensates \cite{Trombettoni}.  Nonequilibrium discrete systems are drawing considerable attention both from fundamental as well as applied points of view. In particular, spatial localization of light in discrete photonic lattices has been reported \cite{Lederer2003,Fleischer2003,Iwanow2004}, including complex confinement of light  such as random-phase solitons \cite{Cohen2005,Buljan2004}. In free propagation, the spatial confinement is attributed to the balance between the discrete diffraction and the nonlinearity. However, when dealing with coupled microresonators, the dissipation of energy due to mirrors should be compensated by optical injection. This second balance renders discrete  dissipative solitons more robust   \cite{Egorov,Egorov2005,Egorov2013}.

Generally speaking, when a system exhibits a simultaneous coexistence between coherence and incoherence behavior in coupled oscillators, the resulting phenomenon is called {\it chimera states} \cite{Strogatz}. Initially, this phenomenon was reported in the context 
of nonlocally  coupled phase oscillators \cite{Strogatz,Kuramoto}, and extended later on to {locally} coupling oscillators \cite{chimera2016, chimera2018}.  In optical systems, experimental observations of chimera states have been reported using  an optoelectronic delayed feedback setup \cite{Larger}, 
laser {diodes} coupled to a nonlinear saturable absorber \cite{Viktorov},
and laser {diodes} subjected to a coherent polarization \cite{Sciamanna}.
Recently, one-dimensional optical chimera states have been predicted in an array of coupled Kerr resonators \cite{chimera2017}.   However, to   the best {of} our knowledge  chimera states in two spatial dimensions have  received limited  attention \cite{Aghdami}.

This paper aims to investigate the formation of two-dimensional optical chimera states in an array of coupled waveguide resonators. This phenomenon occurs in a regime where a coupled waveguide resonators exhibit a coexistence between a coherent and incoherent emission. These discrete structures consist of a localized complex domain embedded in a stable homogenous background. To account for 2D optical chimera states, we use a discrete version of the two-dimensional Lugiato-Lefever equation \cite{Lugiato}.  Based on this model, we show that, depending on the initial condition; this system  can support a family of two-dimensional optical chimera states. Lyapunov exponents and Yorke-Kaplan dimension allow characterizing these structures. Chimera states correspond to localized spatiotemporal chaos. In the Lugiato-Lefever equation, the optical chimera states are excluded. Indeed, this dynamical behavior is {a} peculiarity of discrete systems.

\begin{figure}[t]
\centering\includegraphics[width=8 cm,height=7cm]{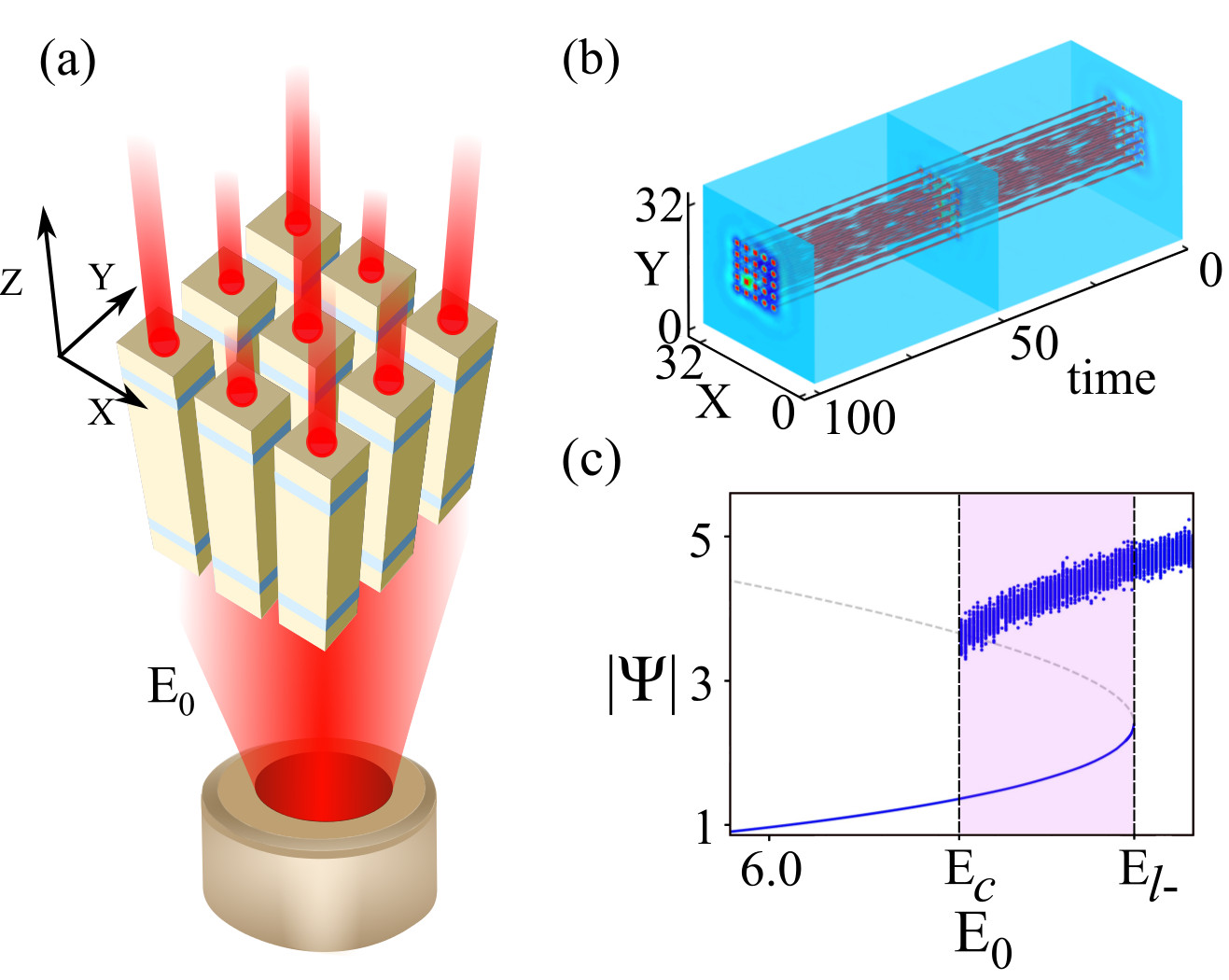}
\caption{Optical chimera states in a two-dimensional array of  coupled microresonator. Parameters are $E_0=4.22$,  $\Delta=-0.506$, and $\kappa=1.876$. 
 (a) Schematic representation of a two-dimensional array of coupled-waveguide resonators driven by an external electrical field of intensity 
 $E_0$. 
 (b) Spatiotemporal evolution of the maximum iso-surface   amplitude of each interacting cavity.%ies. 
 (c) Bifurcation diagram of model Eq.~(\ref{Eq1-DiscreteDDNLS}). The total intracavity amplitude $\left| \Psi \right|$ as function of the pump  amplitude  $E_0$. 
 The solid and dashed lines describe the total
intracavity intensity of homogeneous steady states. The blue cloud of points shows %Points cloud (blue) stands for 
the extreme values of 
the total intracavity intensity of the  spatiotemporal chaotic state. The colored region accounts for the coexistence range  ($E_c=6.68 <E<E_{l_-}= 7.29$).  2D optical 
chimera states are observed inside this interval. }
\label{Fig1-chimeraSetup}
\end{figure}

\section{Array of driven coupled  waveguide resonators: 2D discrete Lugiato-Lefever model}

Let us consider a two-dimensional square array of coupled waveguide resonators subject to a coherent monochromatic beam. Figure~\ref{Eq1-DiscreteDDNLS} shows a schematic representation of  the driven square lattice. 
 Each resonator is composed of a waveguide filled
by a Kerr media, with dielectric mirrors {at}
%in 
the end-faces. Indeed, this system
corresponds to a  {lattice of waveguide microcavities.} %waveguide microcavities lattice.
This  device can be described by the discrete Lugiato-Lefever model \cite{Egorov2005,Egorov2013}. Note that this prototype model of driven coupled oscillators has been more studied in the one-dimensional configuration. Assuming that the coupling between waveguide-resonators is small in comparison with the cavity size, the intracavity field satisfies
\begin{eqnarray}
\partial_T\Psi_{n,m} &=& E_0 - (1+i\Delta)\Psi_{n,m} - i |\Psi_{n,m}|^2\Psi_{n,m}\nonumber\\
&&- i\kappa\left(\Psi_{n+1,m} +\Psi_{n-1,m} + \Psi_{n,m+1}+\Psi_{n,m-1}\right),
 \label{Eq1-DiscreteDDNLS}
\end{eqnarray}
where  $\Psi_{n,m}(T)$ is {a} slowly varying envelope  of the electric field circulating in $(n,m)$-coupled resonators. Indices $n$ ($x$-axis) and $m$  ($y$-axis) denote the transverse coordinates of the cavities.
The detuning parameter 
$\Delta \equiv \omega-\omega_0$ is proportional to the difference between the resonance frequency $\omega_0$ of the cavity and the driving %en
 field frequency $\omega$.  $\kappa$ 
 characterizes the coupling strength between the cavities. 
The time $t = T \tau_{ph}$ is measured in the photon
lifetime unit $\tau_{ph}$. 
The driving field  intensity is denoted by $E_0$.
The continuous counterpart of model Eq.~(\ref{Eq1-DiscreteDDNLS})  was used to describe  Kerr optical frequency combs  (see the special issue \cite{LLEreview} and references therein).

In the continuous limit, for $\Delta> \sqrt{3}$ ($\Delta< \sqrt{3})$, the transmitted intensity as a function of 
the input intensity $E_0^2$ is
bistable (monostable). The homogeneous steady state undergoes a modulational  
instability at $E_0^2=E_{0c}^2\equiv1+(1-\Delta)^2$ and $|\Psi_c|^2=1$. At this bifurcation point, the critical wavelength  is $\Lambda_c^2 = [2 \pi  |\kappa| / (2-\Delta)]^{1/2}$. It has been shown that, for large injected intensity, the system exhibits a spatiotemporal chaos \cite{Liuu2017}.
These dynamic behaviors are persistent when one considers the respective discrete system \cite{chimera2017}.
In this type of systems, the {discreteness} (Peierls-Nabarro potential) allows the confinement of light. Hence{\color{blue},}
the prerequisite  condition for the formation of two-dimensional chimera states is the  coexistence 
of a coherent %ce 
(homogeneous state) and an incoherent %ce
 state (spatiotemporal chaos) in a discrete system.
Figure~\ref{Eq1-DiscreteDDNLS}b displays a typical 2D chimera state in the bistability region, 
$E_{c}<E_{0}<E_{l-}$. 
 Chimera states are classified  by
 the notation $n \times  m$, which depicts the number of cavities that shows
the maximum amplitude.
Numerical simulations  are conducted  using a finite difference
code with a 4th-order Runge-Kutta scheme and Neumann boundary conditions.
Contrary to the continuous limit, the 2D chimera states neither grow in spite of available free space in the transverse plane nor shrink in spite of weak coupling between resonators. 
Figure~\ref{Eq1-DiscreteDDNLS}c shows the bifurcation diagram of model 
Eq.~(\ref{Eq1-DiscreteDDNLS}). We plot the maximum values of the normalized total intracavity  amplitude  $|\Psi|$ as a function the injected field amplitude $E_0$. 
 The normalized total amplitude of the intracavity field is defined as, $|\Psi(t)| = \sqrt{\sum_{i,j=1}^N|\Psi_{i,j}(t)|^2}/N^2$ with $N^2$ is the total number of coupled cavities in the lattice. 
 
 \begin{figure}[b]
\centering\includegraphics[width=8.5cm]{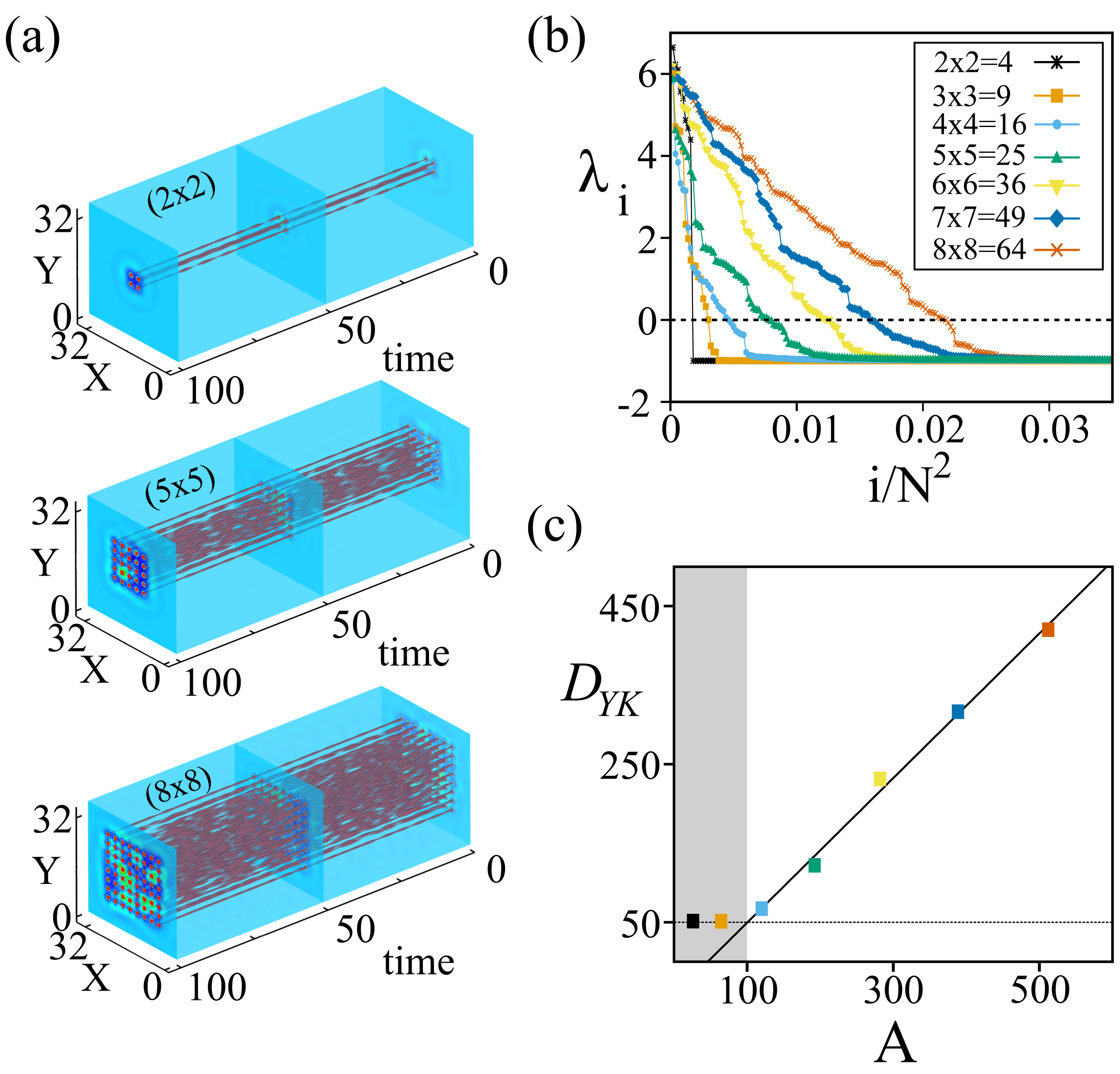}
\caption{Family of  two-dimensional optical chimera states of model Eq.~(\ref{Eq1-DiscreteDDNLS}) with the same parameters {as} %of Fig.~\ref{Fig1-chimeraSetup}.
(a) Spatiotemporal 
diagrams of $2\times2$, $5\times 5$, and $8\times8$ optical chimera states.
 The product $n \times  m$ accounts for the number of cavities that shows
the maximum amplitude.
(b) Lyapunov spectra of different  2D optical chimera states obtained from Eq.~(\ref{Eq1-DiscreteDDNLS}).  
$\{ \lambda_i \}$ denotes the $i$-Lyapunov exponent, $i=\{1,\cdots,N\}$, and   $N$ accounts for 
the total number of cavities. Each curve corresponds to the Lyapunov spectrum of the respective $n \times m$ chimera states. 
(c) Yorke-Kaplan dimension of the spatiotemporal chaotic solution as function of $A$ parameter.
This parameter accounts for the average number 
of microcavities in the incoherence domain.
}
\label{Fig2-Familychimera}
\end{figure}
 
 For small $E_0$, only {a} homogeneous steady state exists as a stable solution. In this case, all cavities have the same intracavity field amplitude. Increasing the input parameter  up to $E_0\geqslant E_{l-}$, the homogeneous steady state suffers a saddle-node instability. The system develops the emergence of spatiotemporal chaos \cite{Liuu2017}. 
 Further increasing  $E_0$, the complex dynamics keeps up. When decreasing $E_0$, the spatiotemporal complex dynamics perseveres until  $E_0$ reaches $E_c$ (see Fig.~\ref{Eq1-DiscreteDDNLS}c). For $E_0 <E_c$, the homogeneous steady state is the only extended stationary 
{ equilibrium}. Indeed, the system presents a subcritical bifurcation at $E_0=E_c$.  The coexistence
  is prerequisite for the  formation scenario of chimera states we have previously proposed in 1D \cite{chimera2016}. 

 The first finding is that the family 
 of chimera states generated in the transverse section 
of the intracavity field is much {more} diverse  
than one-dimensional case, thanks to the large variety of 2D geometrical {plots}. %charts. 
However, for the sake of simplicity, we limit our analysis to chimera states with incoherent domains forming a square as depicted in  Figure~\ref{Fig2-Familychimera}a. 
These chimera states are characterized by spatial 
confinement of large temporal fluctuations (see video in supplement).

\section{Characterization of 2D optical chimera states}
In dynamical systems theory, Lyapunov exponents constitute the most 
adequate tool to characterize  the 
nature of complex spatiotemporal dynamics described above. These exponents provide information
about  sensitivity to the initial conditions, fluctuations, and complexity of solutions \cite{Pikovsky}. 
Low dimensional and spatiotemporal chaos are characterized by positive Lyapunov exponents.
 These exponents can be computed from the method proposed {by Skokos} \cite{Skokos}.  
 The set of Lyapunov  exponents constitutes
 the Lyapunov spectrum $\{ \lambda_i\}$ with $i=\{1,2,\cdots, N^2\}$, $\lambda_i \leqslant \lambda_j$,
and $i \leqslant j$. 
Low-dimensional chaos possesses a discrete Lyapunov spectrum, while spatiotemporal chaos has a continuous one.  
 Figure~\ref{Fig2-Familychimera}b shows Lyapunov spectra of different optical chimera states.
 From this {plots,}%chart, 
we see that positive Lyapunov exponents increase with the size 
 of chimera states.  Hence, the complexity of these localized solutions increase with chimera states
 size.

In addition, from Lyapunov spectrum we can compute the Yorke-Kaplan dimension defined by
$D_{YK} = p + \sum_{i=1}^p \lambda_i/|\lambda_{p+1}|$, 
where $p$ is the {largest} integer for which $\lambda_1 +\dots + \lambda_p > 0$.
 Figure~\ref{Fig2-Familychimera}c shows the Yorke-Kaplan
 dimension of different chimera solutions. 
 From small values of $A $, the Yorke-Kaplan dimension remains constant, where
 $A$ denotes the average number of microcavities in the incoherence domains. 
 As $A$ is increased the Yorke-Kaplan dimension {grows.} %growth. 
This feature is the manifestation 
 of extensive property of this dynamical dimension \cite{Pikovsky}, 
 indicating that 2D optical chimera states 
 belong to the class of spatiotemporal chaos.

\begin{figure}[h]
\centering\includegraphics[width=8.5 cm]{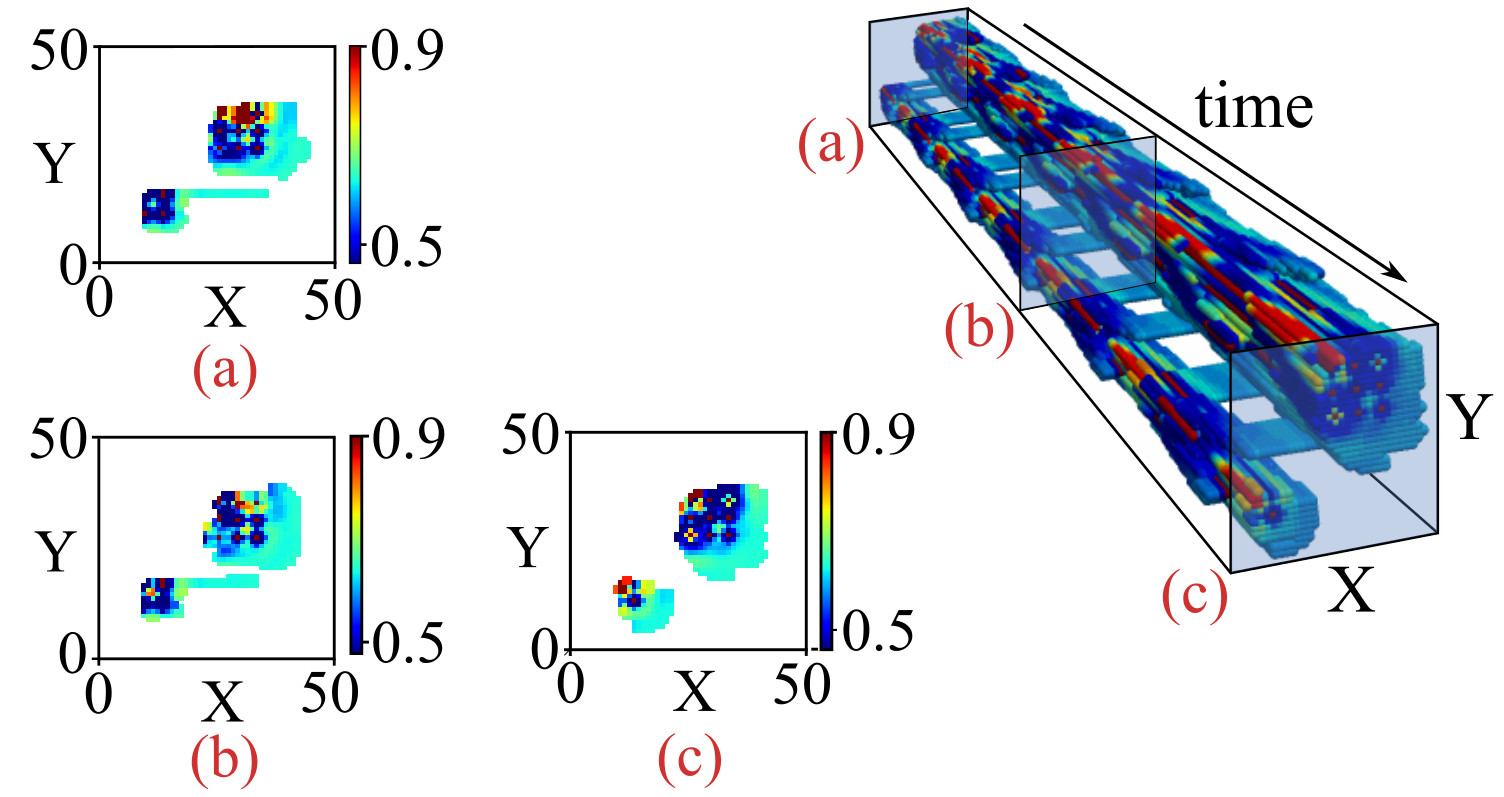}
\caption{2D chimera states exist together. 
Spatiotemporal diagram of 2D chimera states 
in an array of coupled waveguide-resonators cavities. 
The color bar stands for the intracavity intensity field.
Insets (a), (b), and (c) account for the cross section at different time. }
\label{Fig4-Coexistence}
\end{figure}

 \begin{figure}[h]
\centering\includegraphics[width=8cm]{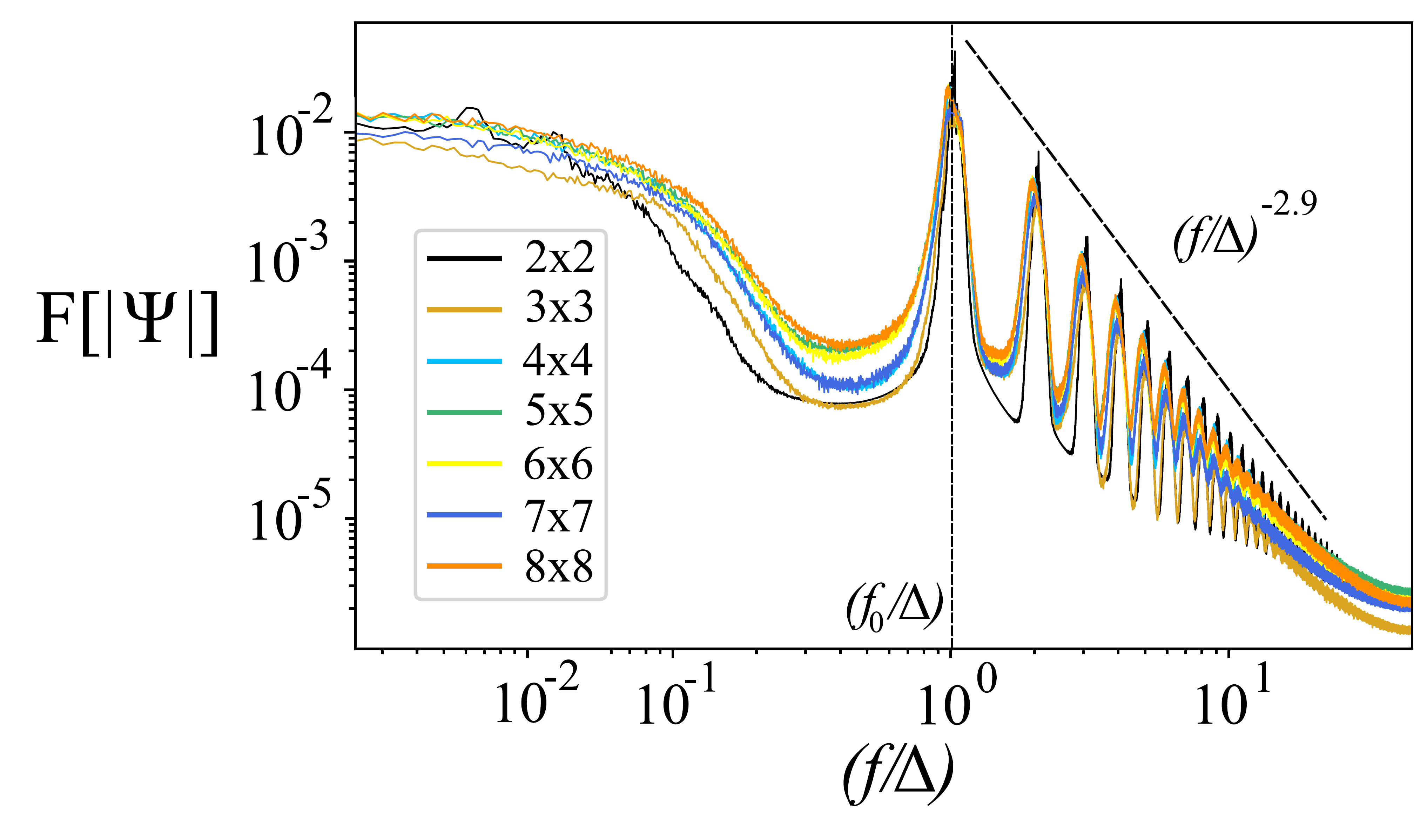}
\caption{Power spectrum $F[|\Psi|]$ of a single waveguide-resonator cavity as function of the frequency $f$
and the detuning parameter $\Delta$. 
At high frequencies, the power spectrum shows 
a power law $f^{-2.9}$ which is a signature of turbulence-like dynamics.}
\label{Fig3-FourierAnalysis}
\end{figure}

 Fourier analysis is used {to} %for 
further characterize the underlying dynamics of
 chimera states. To perform this analysis, we have the spectral density of the signal recorded at the location of one of the largest local maxima in the incoherent domain. Figure \ref{Fig3-FourierAnalysis} shows {the}
 resulting %the 
power spectrum for different chimera states.  The shape of the power spectrum is not affected by the size of the incoherence domain.
The power spectrum has a dominant peak at the value  of the detuning parameter. 
 For high frequencies, the power spectrum presents a power $f^n$ where $n= -2.932$, which is a signature of 
 turbulence-like behavior \cite{Frisch}.

Finally,  numerical simulations of model Eq.~(\ref{Eq1-DiscreteDDNLS}) shows evidence of the coexistence 
between dissimilar chimera states simultaneously in different spatial locations in the transverse plane.
An example of such a behavior is shown in Fig.~\ref{Fig4-Coexistence}, where 
 $2\times 2$ and $3\times 3$ optical chimera
states exist together. Insets account for the cross section at different times.
The 2D spatiotemporal diagram suggest that the chimera states interact weakly.

\section{conclusion}
We have shown evidence of two-dimensional optical chimera states 
in a driven array of locally-coupled passive Kerr optical resonators. Adequate initial conditions have been used to generate a family of these solutions. The main characteristic of these solutions  is spatial confinement of
light in the transverse plane involving complex multi-peaks dynamics. 
Besides, we have shown that these solutions can coexist together. 
The 2D chimera states are inherent  to the discrete nature of the system. 
Indeed, in the continuous limit,  these states are unstable.
We have characterized these solutions by computing Lyapunov spectra, York-Kaplan dimensions, and power spectrum. We have {showed}
%revealed 
that the 2D optical chimera states belong to the class of spatiotemporal chaos and turbulence like behaviors. The prerequisite condition for their formation requires a
bistable behavior between homogeneous background and spatiotemporal chaos.
This condition is rather general, and therefore,  this prediction is important for the identification and understanding of the various complex spatiotemporal behaviors observed in practical systems. 

\section*{supplementary material}
See supplement for supporting contents

\begin{acknowledgments}
M.G.C. and M.A.F. acknowledge the financial support of Millennium Institute for Research in Opticsan  Fondecyt 1180903.  M.T. is a Research Director of the Fonds National de la Recherche Scientifique (Belgium).
\end{acknowledgments}


\nocite{*}
\begin{thebibliography}{1}
\newcommand{\enquote}[1]{``#1''}



\bibitem{Joseph1988} D.~N.~Christodoulides and R.~I.~Joseph, Opt.~Lett.~\textbf{13}, 794 (1988).

\bibitem{Davydov} A.~S.~Davydov and N.~I.~Kislukha, 
Phys. Status Solidi B \textbf{59} 465-470 (1973).

\bibitem{Su1979} W.~P.~ Su, J.~R.~Schieffer and A.~J.~Heeger, Phys. Rev. Lett. \textbf{42},1968-1971 (1979).

\bibitem{Trombettoni} A.~Trombettoni and A.~Smerzi, Phys.  Rev. Lett. \textbf{84}, 5435-5438 (2001).

\bibitem{Lederer2003} D.~N.~Christodoulides, F.~Lederer, and Y.~Silberberg, Nature \textbf{424}, 817 (2003)
;R.~Morandotti, U.~Peschel, J.~S.~Aitchison, H.~S.~Eisenberg, and Y.~Silberberg, Phys.~Rev.~Lett.~ \textbf{81} 3383


\bibitem{Fleischer2003} Fleischer, J. W., Segev, M., Efremidis, N. K., and Christodoulides, D. N. (2003). Nature, 422(6928), 147.; N.~K.~Efremidis, S.~Sears, D.~N.~Christodoulides, 
J.~W.~Fleischer, and M.~Segev,
 Phys.~Rev.~E \textbf{66} 046602 (2002); D.~Neshev, E.~Ostrovaskaya, Y.~Kivshar, and W.~Krolikowski, Opt.~Lett.~\textbf{28}, 710 (2003)


\bibitem{Iwanow2004}A.~Fratalocchi, G.~Assanto, K.~A.~Brzdakiewicz, and M.~A.~Karpierz, Opt.~Lett.~\textbf{29},1530 (2004);
 Y.~V.~Kartashov, V.~A.~Vysloukh, Opt.~Lett.~\textbf{30}, 637 (2005);
 H.~Martin, E.~D.~Eugenieva, Z.~Chen, and D.~N.~Christodoulides, Phys.~Rev.~Lett.~\textbf{92} 123902 (2004);
 B.~A.~Malomed and P.~G.~Kevrekidis, Phys.~Rev.~E \textbf{64}, 026601 (2001)


\bibitem{Cohen2005}O.~Cohen, G.~Bartal, H.~Buljan, T.~Carmon, J.~W.~Fleischer, M.~Segev, and D.~N.~Christodoulides, Nature \textbf{433}, 500 (2005)

\bibitem{Buljan2004} H.~Buljan , O.~Cohen, J.~W.~Fleischer, T.~Schwartz, M.~Segev, Z.~H.~Musslimani, N.~K.~Efremidis,
and D.~N.~Christodoulides,  Phys. Rev. Lett., {\bf 92}, 223901  (2004).


\bibitem{Egorov} U.~Peschel, O.~Egorov, Lederer,  Opt. Lett. \textbf{29} 1909 (2004).

\bibitem{Egorov2005}O.~Egorov,U.~Peschel, and F.~Lederer, Phys.~Rev.~E \textbf{71}, 056612 (2005).
\bibitem{Egorov2013}O.~A.~Egorov and F.~Lederer, Opt.~Lett.~\textbf{38}, 1010 (2013).


\bibitem{Strogatz} D.~M.~Abrams and S.~H.~Strogatz, Phys. ~Rev.~Lett.~\textbf{93}, 174102 (2004).

\bibitem{Kuramoto} Y.~Kuramoto and D.~Battogtokh, Nonlinear Phenom.~ Complex Syst.~\textbf{5}, 380 (2002)

\bibitem{chimera2016} M.~G.~Clerc, S.~Coulibaly, M.~Ferr\'e, M.~A.~Garc\'ia-\~Nustes, and R.~G.~Rojas, Phys. Rev. E \textbf{93}, 052204 (2016).

\bibitem{chimera2018} M.G. Clerc, S. Coulibaly, M.A. Ferr\'e, and R.G. Rojas, 
Chaos, {\bf 28}, 083126 (2018).

\bibitem{Larger}  L. Larger, B. Penkovsky,  and Y. Maistrenko,   
Nat. Comm.  {\bf 6}, 7752 (2015).

\bibitem{Viktorov}  E. A. Viktorov, T. Habruseva, S. P. Hegarty, G. Huyet, and B. Kelleher, Phys. Rev. Lett. 112, 224101 (2014).

\bibitem{Sciamanna}  C.-H. Uy, L. Weicker,   D. Rontani, and M. Sciamanna, APL Photon. {\bf 4}, 056104 (2019).

\bibitem{chimera2017} M.~G.~Clerc, M.~A. Ferr\'e, S.~Coulibaly, R.~G.~Rojas, and M.~Tlidi,  Opt. Lett. \textbf{42}, 2906 (2017).



\bibitem{Aghdami} K. M. Aghdami, M., Golshani, and R. Kheradmand,  
IEEE Photonics journal, {\bf 4}, 1147 (2012).



\bibitem{Lugiato} L.~A.~Lugiato and R.~Lefever, 
Phys.~Rev.~Lett. \textbf{58},2209 (1987). 

\bibitem{LLEreview} Y.K. Chembo, D. Gomila, M. Tlidi, and C.R. Menyuk,
Eur. Phys. J. D {\bf 71}, 299 (2017).


\bibitem{Liuu2017} Z. Liu, M. Ouali, S. Coulibaly, M.G. Clerc, M. Taki, and M. Tlidi,   
Opt. lett., {\bf 42}, 1063 (2017).

\bibitem{Pikovsky}A. Pikovsky and A. Politi, Lyapunov Exponents: A Tool to Explore Complex Dynamics (Cambridge University, 2016).

\bibitem{Skokos}C. H. Skokos, \enquote{ The Lyapunov characteristic exponents and their computation}, in {\it Dynamics of Small Solar System Bodies and Exoplanets} (Springer, 2010), pp. 63–135.

\bibitem{Frisch}U. Frisch, and A. N. Kolmogorov, Turbulence: the legacy of AN Kolmogorov (Cambridge university press, 1995).

%\bibitem{Reinisch} {\color{red}R.~Reinisch, E.~Popov, and M.~Neviere, \enquote{Second-harmonic-generation-induced optical bistability in prism or grating couplers}, Opt.Lett. \textbf{20}, 854 (1995).}






\end{thebibliography}
\end{document}